\documentclass[12pt]{article}

\usepackage{amsmath,amssymb,graphicx, float} 
\usepackage{epsf}
\usepackage{pstricks}
\usepackage[sort&compress,numbers]{natbib}
\usepackage{hyperref}
\usepackage{ctable}
\usepackage{multirow}
\usepackage{enumerate}

\def\beq{\begin{equation}}
\def\eq{\end{equation}}
\def\eeq{\end{equation}}

\def\centeron#1#2{{\setbox0=\hbox{#1}\setbox1=\hbox{#2}\ifdim
\wd1>\wd0\kern.5\wd1\kern-.5\wd0\fi
\copy0\kern-.5\wd0\kern-.5\wd1\copy1\ifdim\wd0>\wd1
\kern.5\wd0\kern-.5\wd1\fi}}
\def\ltap{\;\centeron{\raise.35ex\hbox{$<$}}{\lower.65ex\hbox{$\sim$}}\;}
\def\gtap{\;\centeron{\raise.35ex\hbox{$>$}}{\lower.65ex\hbox{$\sim$}}\;}

\def\D0{D$\O$}

\begin{document}

\begin{titlepage}

\begin{center}
\vspace*{-1cm}

\hfill RU-NHETC-2011-07 \\
\vskip 1.3in
{\LARGE \bf Signatures of Resonant Super-Partner} \\
\vspace{.15in}
{\LARGE \bf Production with Charged-Current Decays}
\vspace{.15in}

\vskip 0.65in
{\large Can Kilic}
{\rm and} \large {Scott Thomas}
\vskip 0.25in

{ \normalsize \it New High Energy Theory Center \\
Department of Physics \\
{Rutgers University} \\
\vspace{-.05in}
Piscataway, NJ 08854}

\vskip 0.75in

\end{center}

\baselineskip=16pt

\begin{abstract}

Hadron collider signatures of new physics are investigated in which a primary resonance is produced that decays to a secondary resonance by emitting a W-boson, with the secondary resonance decaying to two jets. This topology can arise in supersymmetric theories with $R$-parity violation where the lightest supersymmetric particles are either a pair of squarks, or a slepton - sneutrino pair.  The resulting signal can have a cross section consistent with the $Wjj$ observation reported by the CDF collaboration, while remaining consistent with earlier constraints.  Other observables that can be used to confirm this scenario   include a significant charge asymmetry in the same channel at the LHC. With strongly interacting resonances such as squarks, pair production topologies additionally give rise to 4 jet and W$^+$W$^-$+ 4 jet signatures, each with two equal-mass dijet resonances within the 4 jets.

\noindent

\end{abstract}

\end{titlepage}

\baselineskip=17pt

\newpage





\section{Introduction}

In the overwhelming majority of the literature on supersymmetric extensions of the standard model, $R$-parity is imposed as an ad-hoc symmetry to avoid phenomena such as proton decay which has been very strongly constrained by experiment. However, the proton can remain exactly stable even in the presence of a restricted set of $R$-parity violating interactions. In fact, one needs to break both baryon and lepton number for the proton to decay, because the proton is the lightest fermion that carries zero lepton number but nonzero baryon number. Various low energy constraints on $R$-parity violating couplings have been studied in depth (see \cite{RPVreview} and references therein), however certain aspects of collider phenomenology in the presence of $R$-parity violating couplings have not received the same amount of attention compared to the case of exact $R$-parity \cite{SUSYcolliderpheno}. In particular, there are two major differences between $R$-parity conserving and $R$-parity violating supersymmetric models that we wish to focus on in this paper that relate to the signatures at hadron colliders.

Firstly, $R$-parity violation affects production mechanisms by allowing for supersymmetric partner particles to be resonantly produced \cite{Dimopoulos:1988fr,Dreiner:1991pe}. Secondly, the lightest supersymmetric particle need no longer be stable, and can decay promptly. Since most experimental searches for supersymmetry focus on the presence of missing transverse energy carried off by a neutral lightest superpartner, this requires a major change in search strategies. Naturally, in order to account for dark matter, a supersymmetric theory with $R$-parity violation would need to have additional matter content with a symmetry that can stabilize the lightest particle in the dark sector. Finally, the fact that the lightest supersymmetric particle is no longer stable allows for spectra that are phenomenologically disfavored in models with exact $R$-parity and a colored superpartner can be the lightest one in the spectrum\cite{Sarid:1999zx}. In the rest of this paper we wish to study some of the  phenomenological consequences associated with  $R$-parity violation with relatively light scalar superpartners, such as the stop.

UV motivated models of $R$-parity violation have been studied in the context of mSUGRA and grand unified theories \cite{Sakai:1981pk,Hall:1983id,Brahm:1989iy,Hempfling:1995wj,Smirnov:1995ey,Tamvakis:1996xf,Tamvakis:1996dk,Barbieri:1997zn,Giudice:1997wb,Dreiner:1997uz,Diaz:1997xc,Allanach:2003eb}. There also exist motivated models of new physics that are qualitatively different than the MSSM, in which $R$-parity is violated. An example is provided by the setup of ref.~\cite{Sundrum:2009gv}, where supersymmetry is broken at high energies but reemerges as an accidental symmetry at low energies, and $R$-parity violation is
unavoidable \cite{Sundrum_private}.

Very recently, the CDF collaboration has announced an interesting observation of a resonance in dijets produced in association with a W boson \cite{CDFanomaly}. The significance of this observation is $3.2\sigma$ using data with $4.3~{\rm fb}^{-1}$ of integrated luminosity, and the mean value for the resonant dijet mass distribution is around 145 GeV. In this paper we will investigate the possibility that the CDF observation arises from the production and decay of new particles, and explore how this final state can arise in models of supersymmetry with $R$-parity violation. The topology we consider is characterized by the production of a primary resonance that undergoes a charged-current transition to a secondary resonance by emitting a W-boson, with the secondary resonance subsequently decaying to two jets. For a supersymmetric realization of this topology, we will concentrate on the possibility of  squarks or sleptons and sneutrino being the lightest (s)particles in the spectrum. We give a representative model in which the primary resonance is a bottom squark (sbottom), and the secondary resonance a top squark (stop), the production and decay of which occur through $R$-parity violating interactions respectively. We will show that only one or two couplings beyond the MSSM are needed for this topology to be consistent with the CDF observation. We will also show that the required couplings can have large enough values without causing conflict with any existing constraints, and we will furthermore argue that the $Wjj$ channel is in fact naturally the first place for the new physics to be observed in this scenario.

In the next section we present our sbottom-stop model, estimate the size of the necessary $R$-parity violating couplings and show that these are not in conflict with any existing bounds. We will then go through various production and decay possibilities and argue why the new physics would appear first in the $Wjj$ channel, and evaluate the discovery potential in other final states for the Tevatron as well as the LHC. In section \ref{sec:sleptons} we give an alternative setup with sleptons instead of squarks that can give rise to similar collider signatures and highlight the differences in phenomenology compared to the stop-sbottom model.

It should be emphasized that the most important aspect of the scenarios presented here are the production and decay topologies. While the language of supersymmetry is utilized throughout, the topologies and signatures may be more widely applicable to other frameworks for the underlying physics, including ones with particles of different spin and gauge quantum numbers.

\section{Resonant Squark Production And Decay}
\label{sec:squarks}

The $R$-parity violating terms in the superpotential are typically parameterized as
\begin{equation}
W_{RPV}=\mu_{i}H_{u}L_{i}+\frac{1}{2}\lambda_{ijk}L_{i}L_{j}E^{c}_{k}+\lambda^{'}_{ijk}L_{i}Q_{j}D^{c}_{k}+\frac{1}{2}\lambda^{''}_{ijk}U^{c}_{i}D^{c}_{j}D^{c}_{k}
\end{equation}
where $i,j,k$ are flavor indices. The first three types of terms violate lepton number while the $UDD$ type terms violate baryon number.

Among the available $R$-parity violating couplings, $UDD$ type terms are generically less constrained than $LQD$ or $LLE$ terms, since searches for new physics involving leptons typically have higher sensitivity. The main constraints on $UDD$ type terms come from neutron-antineutron oscillations and flavor violation in the quark sector\cite{RPVreview}. These constraints can be avoided however, as we will outline below, and for certain choices of $ijk$, $\lambda^{''}_{ijk}$ can be of order one. For resonant production we will need at least one coupling to be reasonably large, and this coupling should preferably involve two light quarks in order to give rise to interesting production cross sections.

Note that the $UDD$ type terms are antisymmetric in the indices $j$ and $k$. Therefore if two of the indices are to be 1, the only choices are $\lambda^{''}_{112}$ and $\lambda^{''}_{113}$, both of which are severely constrained from nuclear decays involving two neutrons\cite{RPVreview}. The next best possibility for resonant production is to make use of a strange quark PDF in the proton, and the there are several couplings of this type that are essentially unconstrained. Note however that there are many additional constraints on products of couplings and therefore the safest choice is to turn on as few $R$-parity violating terms as possible. Below, we will outline a model where the stop and sbottom are the main players for the phenomenology, and where only $\lambda^{''}_{123}\sim{\mathcal O}(0.1)$ and $\lambda^{''}_{323}$ is nonzero (but need not be large). At the end of this section we will point to an alternative choice with only $\lambda^{''}_{212}$ turned on and argue that given a similar spectrum, the collider signatures are very similar.

\begin{figure}[t]
\begin{center}
\includegraphics[scale=0.5]{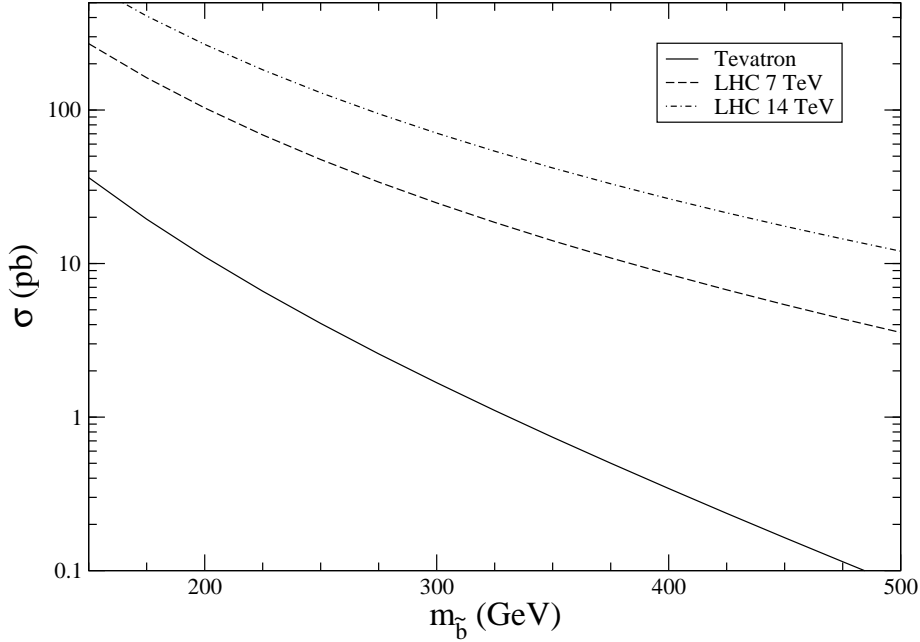}
\end{center}
\caption{Resonant production cross section for sbottom with $\lambda^{''}_{123}=0.1$ as a function of the sbottom mass. For this plot, the mass eigenstate is taken to be pure $\tilde{b}_{R}$.}
\label{fig:RPVmodel}
\end{figure}

\begin{figure}[t]
\begin{center}
\includegraphics[scale=0.5]{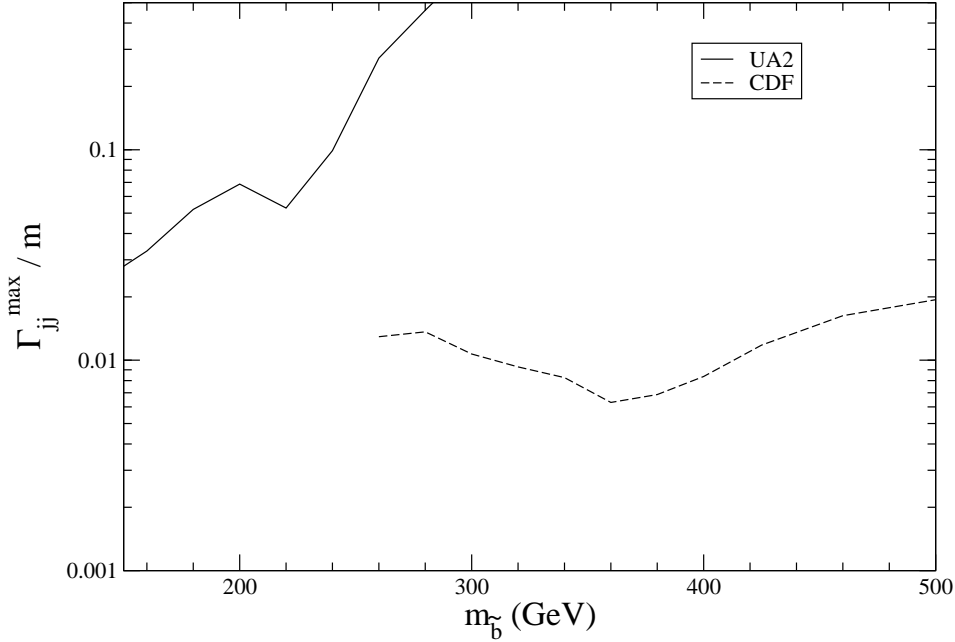}
\end{center}
\caption{The bound on $\Gamma_{jj}$ from the dijet resonance searches at UA2 and CDF. The numbers plotted here are conservative in the sense that we assume perfect acceptance and no suppression from left-right sbottom mixing. In terms of $\lambda^{''}_{123}$, the bound is always above 0.4.}
\label{fig:lambdabound}
\end{figure}

With $\lambda^{''}_{123}$ turned on, sbottoms can be resonantly produced according to $u+s\rightarrow\tilde{b}^{*}_{R}$. The cross section for this process is plotted in figure \ref{fig:RPVmodel} for a value of $\lambda^{''}_{123}=0.1$. If the sbottom happens to be the lightest state, then it will decay back through the $R$-parity violating coupling to a pair of jets. Depending on the sbottom mass, this process is constrained by dijet resonance searches from UA2 \cite{UA2bound} (light sbottoms), at the Tevatron \cite{CDF_dijet} (intermediate mass sbottoms) and recent constraints from the LHC \cite{Aad:2011aj,Khachatryan:2010jd} for heavy sbottoms. In figure \ref{fig:lambdabound} we use the cross section constraints to derive bounds on $\lambda^{''}_{123}$, or equivalently on $\Gamma_{jj}$, the partial width of the sbottom to $u$+$s$. The exact relation between $\Gamma_{jj}$ and $\lambda^{''}_{123}$ is given in equation (\ref{eqn:partialwidths}). In terms of $\lambda^{''}_{123}$ we find that the constraint from resonant production never goes below $\lambda^{''}_{123}=0.4$.

The pair production of sbottoms through QCD has a threshold at twice the sbottom mass, and therefore the cross section for this process is much lower than resonant production. With no other supersymmetric particles lighter than the sbottom, this process would manifest itself in a four jet final state, we will however introduce a different choice of spectrum shortly. It is however interesting to think of the pair production of the lightest colored supersymmetric particle, and we will come back to this point in the next chapter.

The collider prospects for this very minimal (s)particle content are bleak for a value of $\lambda^{''}_{123}$ below the bound. The presence of a neutralino below the mass of the sbottom does not lead to much improvement. The sbottom would decay as $\tilde{b}\rightarrow b+\chi^{0}$, the neutralino then decaying through an off-shell sbottom as $\chi^{0}\rightarrow bjj$, thus populating the $bbjj$ final state, which has a large background. If there are charginos significantly lighter than the sbottom, this can lead to top quarks in the final state, which is potentially a more optimistic scenario, however the large mass gap necessary for decaying to on-shell tops forces the sbottoms to be heavier, reducing the cross section as a function of the lightest superpartner mass.

\begin{figure}[t]
\begin{center}
\includegraphics[scale=0.7]{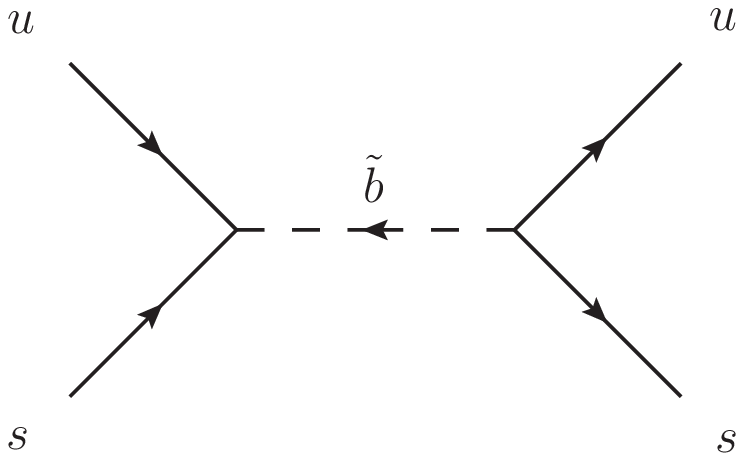}
\includegraphics[scale=0.7]{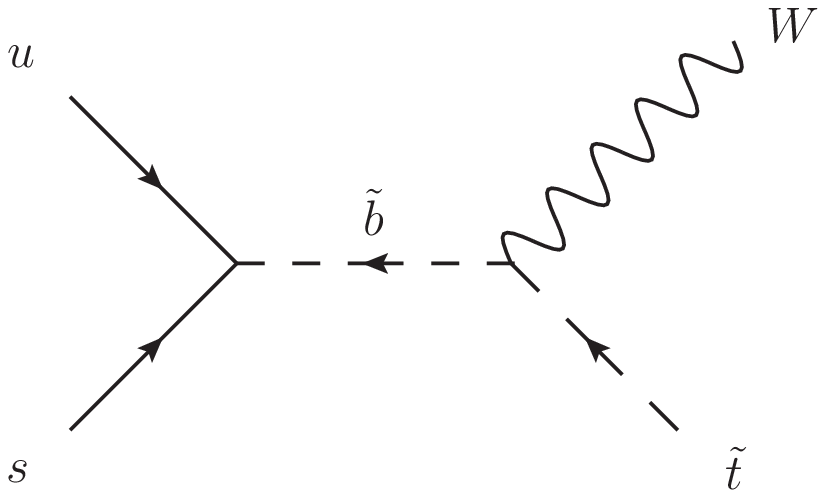}
\end{center}
\caption{Resonant sbottom production and decay modes with a lighter stop in the spectrum.}
\label{fig:sbottommodes}
\end{figure}

\begin{figure}[t]
\begin{center}
\includegraphics[scale=0.5]{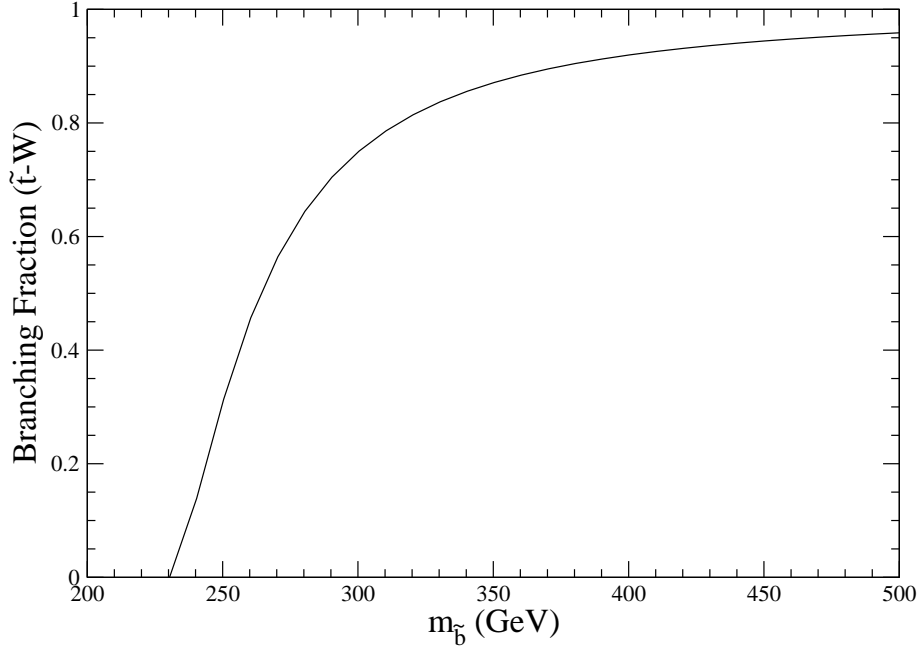}
\end{center}
\caption{The branching fraction of the $\tilde{b}$ to $\tilde{t}$+$W$. For this plot $m_{\tilde{t}}=150$GeV, and $\cos \theta_{\tilde{t}}=\cos \theta_{\tilde{b}}=1/2$.}
\label{fig:sbottomBF}
\end{figure}

A possibility that is much more interesting collider-wise, and the one we wish to consider in the rest of this section is the presence of a stop below the sbottom mass. In this case the sbottom can decay as $\tilde{b}\rightarrow \tilde{t}+W$. This mass spectrum is favored by the RG evolution of the squark masses, as the large Yukawa couplings tend to drive the stop mass lighter. Note that since it is $\tilde{b}_{R}$ that is resonantly produced, for the W channel to dominate there needs to be some mixing between $\tilde{b}_{R}$ and $\tilde{b}_{L}$. As long as the left handed squarks are not decoupled however, the W decay mode can be dominant. The partial widths of the sbottom into dijets and into $\tilde{t}$+$W$ are given by
\begin{eqnarray}
\Gamma( \widetilde{b}_1 \to W^- \widetilde{t}_1) &=&
{g^2 \cos \theta_{\tilde{t}}^2 \cos \theta_{\tilde{b}}^2
~ m_{\widetilde{b}_1}^3 \over 32 \pi m_W^2}
~f^{3/2}(1, m_{\widetilde{t}_1}^2 / m_{\widetilde{b}_1}^2,
 m_W^2 / m_{\widetilde{b}_1}^2 )
  \nonumber\\
\Gamma(  \widetilde{{b}}_1 \to \overline{u} \overline{s} ) &=&
{ |\lambda^{''}_{123}|^2 \sin \theta_{\tilde{b}}^2
~m_{\widetilde{{b}}_1} \over 8 \pi }
\label{eqn:partialwidths}
\end{eqnarray}
where
$f(1,x,y) = 1 - 2(x+y)+(x-y)^2$ is the triangle function and $\cos \theta_{\tilde{b},\tilde{t}}$ denote the mixing angles in the squark sector (for $\cos\theta_{\tilde{b}}=0$, $\tilde{b}_{1}=\tilde{b}_{R}$ and likewise for the stop). In figure \ref{fig:sbottomBF} we plot the branching fraction of $\tilde{b}\rightarrow \tilde{t}+W$ for $\cos \theta_{\tilde{b},\tilde{t}}=1/2$ as a function of the sbottom mass, taking $m_{\tilde{t}}=150$GeV.

\begin{figure}[t]
\begin{center}
\includegraphics[scale=0.85]{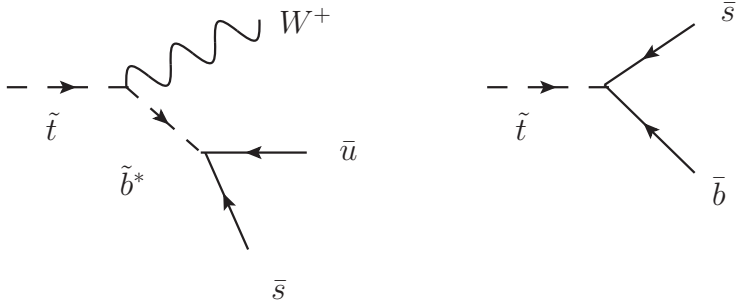}
\end{center}
\caption{Decay modes for the stop.}
\label{fig:stopdecays}
\end{figure}

With only $\lambda^{''}_{123}$ nonzero, the stop will have a three-body decay through on off-shell sbottom. Note however that the stop can also have two-body decays if there are additional nonzero $R$-parity violating couplings, specifically a nonzero value of $\lambda^{''}_{3ij}$. Note however that neutron-antineutron mixing constrains $\lambda^{''}_{312}$ and $\lambda^{''}_{313}$ to be of order $10^{-3}$ while the constraints are much weaker on $\lambda^{''}_{323}$\cite{RPVreview}. Note that the product of $\lambda^{''}_{123}$ and $\lambda^{''}_{323}$ is also not further constrained, and that stop production from a small value of $\lambda^{''}_{323}$ is unobservable due to the PDFs for a $s$-$b$ initial state. The decay modes of the stop are illustrated in figure \ref{fig:stopdecays}.

With $\lambda^{''}_{323}$ turned on, the two-body decay of the stop into jets can dominate, and the first signal to be seen would appear in the $p\bar{p}\rightarrow\tilde{t}W\rightarrow Wjj$ channel. This is the channel where CDF recently reported observing an interesting signal\cite{CDFanomaly}. For this specific choice of couplings, one of the jets from the decay of the stop would be a b-jet. Currently the CDF analysis states that the heavy flavor content of the excess region is consistent with the sideband. However the analysis does not provide a quantitative measure for whether the existence of a single b-jet is disfavored. As we will describe in section \ref{sec:scharm}, the same topology can be realized with a different choice of nonzero $\lambda^{''}_{ijk}$ such that the final state has no heavy flavor jets. The CDF analysis reports no significant deviation from the Standard Model expectation in an analogous channel with a dijet resonance produced in association with a Z-boson. Note that in our minimal setup, this channel is expected to be absent.

In order to study the acceptance of the CDF analysis for the $\tilde{b}\rightarrow W\tilde{t}\rightarrow Wjj$ signal we have performed a Monte Carlo study. It is not trivial to generate Monte Carlo events for this setup due to the completely antisymmetric way that the color indices are contracted in the $R$-parity violating vertex. While certain programs such as HERWIG incorporate aspects of $R$-parity violating physics, this is not adequate for our purposes and we have generated $2\rightarrow 2$ events by using CTEQ parton distribution functions \cite{CTEQ6} and integrating over phase space using a custom-made code. We then exported the events into LHE format \cite{Boos:2001cv} by modifying the color flow to remove the antisymmetric color contractions, which cannot be faithfully represented in the LHE format. While this may introduce subtle changes at the level of non-perturbative QCD effects, we expect that the impact on the hard scattering observables will be negligible. The events were passed through Pythia\cite{PYTHIA} for showering and hadronization, and later through PGS\cite{PGS} for detector effects. We used the standard CDF parameter set in PGS with a cone jet algorithm and a cone size $\Delta R=0.5$.

\begin{figure}[t]
\begin{center}
\includegraphics[scale=0.5]{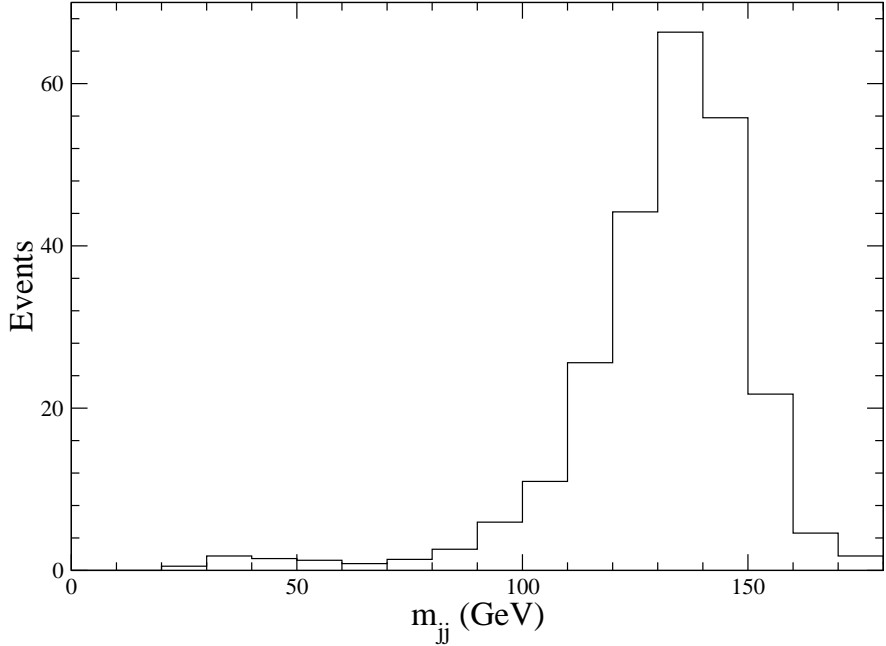}
\end{center}
\caption{The invariant mass distribution of the two jets from the stop decay ($e$ and $\mu$ channels combined) using the same cuts as the CDF analysis, with $m_{\tilde{b}}=300$GeV and $m_{\tilde{t}}=150$GeV. $\lambda^{''}_{123}=0.16$ and $\cos \theta_{\tilde{t}}=\cos \theta_{\tilde{b}}=1/2$ are used in order for the number of events after cuts to agree with the CDF observation with 4.3 fb$^{-1}$ of integrated luminosity.}
\label{fig:mjj_300}
\end{figure}

We analyzed the signal events using cuts designed to mimic the CDF event selection cuts, namely:
\begin{itemize}
\item The presence of an electron (muon) with $E_T$ ($p_T$) $>20~$GeV and $|\eta|<1.0$ and the absence of additional leptons in the event.
\item Missing transverse energy in excess of 25 GeV.
\item The presence of exactly two jets with $E_T>30~$GeV and $|\eta|<2.4$.
\item A minimum $\Delta R$ of 0.52 between the lepton and the nearest jet.
\item A minimum $p_T$ of 40 GeV for the dijet system.
\item A transverse mass in excess of 30 GeV for the lepton-neutrino system, where the neutrino is taken to be the only source of missing transverse energy in the event.
\item An azimuthal angle in excess of 0.4 radians between the missing energy and either jet.
\item $\Delta \eta<2.5$ for the two jets.
\end{itemize}
Taking $m_{\tilde{b}}=300$GeV, $m_{\tilde{t}}=150$GeV and $\cos \theta_{\tilde{t}}=\cos \theta_{\tilde{b}}=1/2$, we find that the number of events observed at CDF in the $e$+$\mu$ channels is consistent with $\lambda^{''}_{123}=0.16$, giving a total production cross section of $3.2$pb, a signal acceptance of $1.6\times 10^{-2}$ which includes a 52\% branching fraction for $\tilde{b}\rightarrow W+\tilde{t}$ (we take the branching fraction of $\tilde{t}\rightarrow jj$ to be 1). This value of $\lambda^{''}_{123}$ is far below the bound from dijet resonance searches. We plot the dijet invariant mass distribution from the stop decay in our sample in figure \ref{fig:mjj_300}.

\begin{figure}[t]
\begin{center}
\includegraphics[scale=0.5]{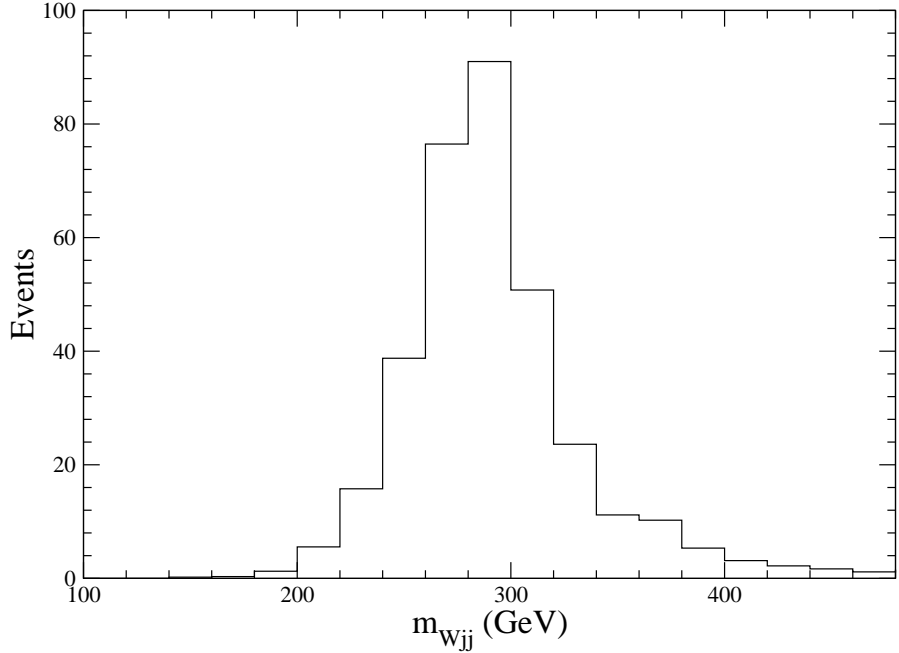}
\end{center}
\caption{The invariant mass of the $Wjj$ system from the sbottom decay ($e$ and $\mu$ channels combined) using the same cuts as the CDF analysis, with $m_{\tilde{b}}=300$GeV and $m_{\tilde{t}}=150$GeV. Both solutions for the neutrino momentum are included in the distribution. $\lambda^{''}_{123}=0.16$ and $\cos \theta_{\tilde{t}}=\cos \theta_{\tilde{b}}=1/2$ are used in order for the number of events after cuts to agree with the CDF observation with 4.3 fb$^{-1}$ of integrated luminosity.}
\label{fig:mwjj_300}
\end{figure}

Another interesting question that is reported to be consistent with the background hypothesis but not ruled out by the CDF analysis is whether a primary resonance can be present in the $Wjj$ system. In the topology considered here, this system will reconstruct the sbottom mass. In figure \ref{fig:mwjj_300} we plot the reconstructed sbottom mass in our model for a true value of $m_{\tilde{b}}=300$GeV using both solutions for the neutrino rapidity in reconstructing the W-mass (and discarding the event if no real solutions can be found). Note that while the stop mass is determined by the position of the observed dijet excess, the sbottom mass is an adjustable parameter in this model, and for lighter sbottom masses, the excess in $m_{Wjj}$ may be difficult to resolve on top of background. A definitive comparison of the primary resonance aspect of this topology with the CDF observation would require a full detector simulation of both signal and backgrounds, as well as the CDF data-driven estimates of background.

\subsubsection*{Additional Collider Signatures}

\begin{figure}[t]
\begin{center}
\includegraphics[scale=0.5]{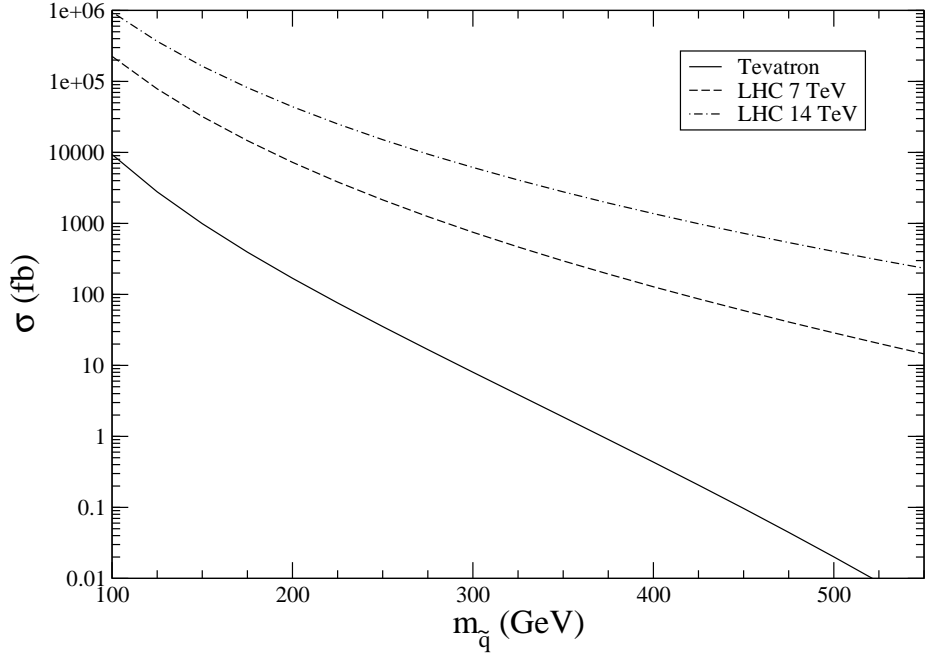}
\end{center}
\caption{Cross section (QCD only) for pair production of a single squark as a function of its mass.}
\label{fig:stoppairxsecn}
\end{figure}

While in our topology the resonant production of any up-type squark is strongly suppressed due to the smallness of the couplings $\lambda^{''}_{k12}$, the pair production of the stop is determined by QCD and can be an interesting production channel. In figure \ref{fig:stoppairxsecn} we show the stop pair production cross section as a function of its mass calculated by using the batch mode of CalcHEP\cite{calchep}. This production channel would manifest itself in the four jet channel, with two pairs of jets reconstructing narrow resonances of the same mass. Ref.~\cite{Choudhury:2005dg} studied this channel with the conclusion that stops with up to $m_{\tilde{t}}=210~{\rm GeV}$ can be discovered. References \cite{Chivukula:1991zk,Dobrescu:2007yp,Kilic:2008pm,Kilic:2008ub,Bai:2010dj} also studied the same event topology at the Tevatron and the LHC for various scenarios of underlying physics, and showed that there is strong discovery potential. Currently no search for new physics has been performed in this channel. Note that for a stop mass of 150GeV, a search in the four jet channel analysis will be difficult at the LHC because of the larger trigger thresholds which will have low efficiency on the signal. The Tevatron has a unique advantage for the case of light stops, and an analysis of existing data could be used to explore this possibility.

\begin{figure}[t]
\begin{center}
\includegraphics[scale=0.5]{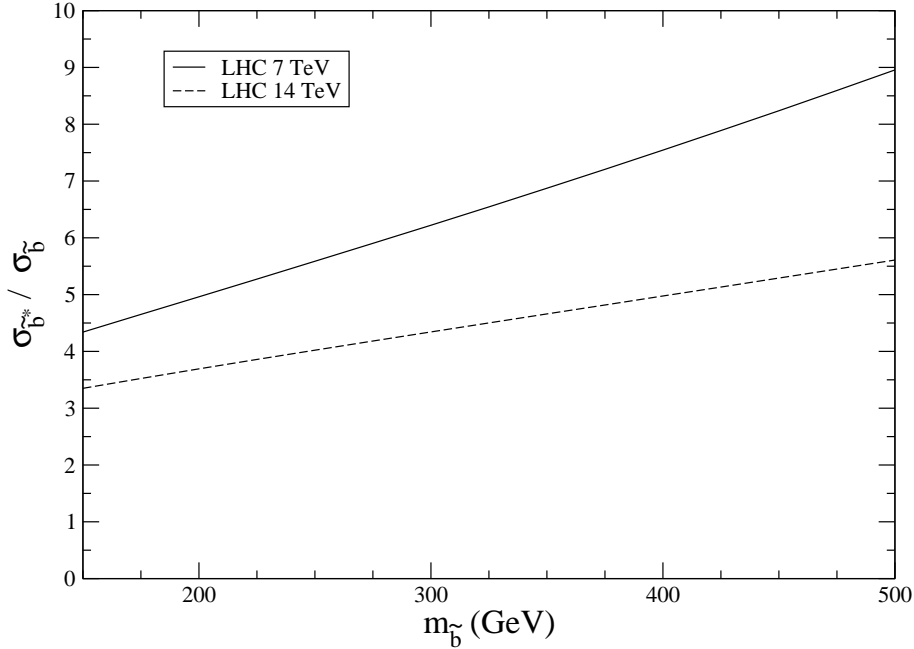}
\end{center}
\caption{Ratio of the anti-sbottom to sbottom resonant production cross sections through the $\lambda^{''}_{123}$ $R$-parity violating
interaction at the LHC.}
\label{fig:chargeasymmetry}
\end{figure}

The initial state for resonant anti-squark production through $R$-parity violating interactions is quark--quark (rather than quark--anti-quark). In the scenario described here, with production through the $\lambda^{''}_{123}$ coupling, one of the initial state quarks is a strange-quark. At the Tevatron with $p$-$\bar{p}$ collisions, as well as at the LHC with $p$-$p$ collisions, the parton distribution functions for the leading resonant production are therefore of the valence--sea type. Aside from the relative center of mass energies, there is no qualitative difference for the overall  production rates between the Tevatron and LHC for such a resonance. However, the relative rates for sbottom versus anti-sbottom are significantly different. At the Tevatron sbottom and anti-sbottom are produced at equal rates because of the charge symmetry of the colliding particles. In contrast, at the LHC, production of anti-sbottom arising from valence--sea quark collisions is enhanced compared with sbottom production arising from sea--sea anti-quark collisions. This leads to a significant charge asymmetry in the $Wjj$ signal at the LHC, as shown in figure \ref{fig:chargeasymmetry}. A charge asymmetry is also present in the W+jets background due to the difference between the $u$ and $d$ valence quark parton distribution functions. However, the W+jets background asymmetry arises predominantly from an order one difference in the relative magnitude of two types of valence--sea quark collisions, whereas the signal asymmetry is much larger since it arises from a difference between valence--sea versus sea--sea collisions. The enhanced charge asymmetry of the $Wjj$ signal may provide an additional handle at the LHC to isolate this supersymmetric scenario for the topology of a primary and secondary resonance with a charged-current transition.

Finally, one can also consider the pair production of the sbottom through QCD. In the scenario with a lighter stop, this channel will go to $W^{+}W^{-}+4j$. While the combinatoric background can be reduced by reconstructing the stops as well as demanding equal sbottom masses, the cross section is small compared to the irreducible standard model backgrounds of $t\bar{t}$+jets and $WW$+jets in the dileptonic decay channel, and the additional background of W+jets for the semileptonic case. As one can see in figure \ref{fig:stoppairxsecn}, even for moderately light sbottom masses, the cross section for this process is too low at the Tevatron. The leptonic branching fraction of $W$ bosons and acceptance effects make it unlikely that a statistically significant excess can be observed at the 7 TeV LHC as well (assuming a total integrated luminosity of $5~{\rm fb}^{-1}$). This channel may become interesting for the 14 TeV LHC with high statistics however.

\subsubsection*{Squark Mixing}

\begin{figure}[t]
\begin{center}
\includegraphics[scale=0.7]{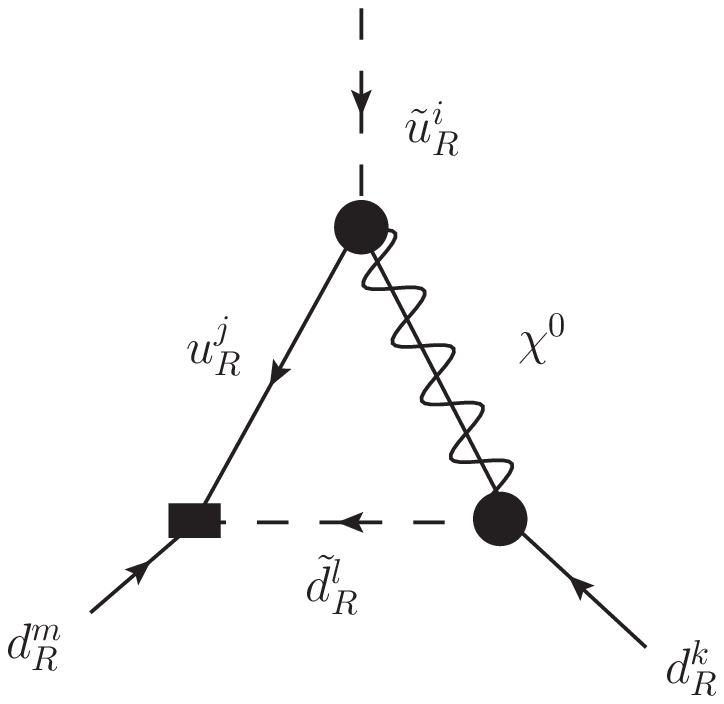}
\end{center}
\caption{How additional $UDD$ type couplings may be induced at 1-loop from existing ones. The black box denotes a $\lambda^{''}$ coupling that is nonzero at tree level and the circles represent off diagonal couplings in flavor space.}
\label{fig:flav}
\end{figure}

We wish to briefly comment on the stability of the choice of nonzero $\lambda^{''}_{ijk}$ that we use. For instance, the 1-loop diagram in figure \ref{fig:flav} shows how an existing $UDD$ coupling may induce additional $R$-parity violating couplings in the presence of neutralinos that couple off-diagonally in flavor space. This can be avoided however if all three generations are nearly-degenerate. For the model we outlined in this paper, one can imagine a mass spectrum such that all three generations are present, with $m_{\tilde{u}_{R}} < m_{\tilde{d}_{R}} < m_{\tilde{q}_{L}}$. The presence of the additional squarks does not significantly change the phenomenology that we discussed. With only $\lambda^{''}_{123}$ and $\lambda^{''}_{323}$ turned on, the $\tilde{u}^{i}_{R}$ are not resonantly produced at observable rates. They are pair produced, enhancing the cross section by a factor of three compared to that of a single stop in the four jet channel. Resonant production of the additional $\tilde{d}^{i}_{R}$ is also suppressed by the PDF's. Therefore the phenomenology is essentially identical to the minimal scenario with only a sbottom and a stop. If the left-handed squarks are light enough to be produced, new decay channels with on-shell Higgs or Z bosons may open up. Other operators induced by a diagram similar to figure \ref{fig:flav} but with a W in the loop rather than a neutralino are nonrenormalizable, and are small for our choice of nonzero $\lambda^{''}$ since they are suppressed by quark masses and CKM angles. We do not consider operators induced at higher loops.

\subsubsection*{An Alternative Choice of Couplings}
\label{sec:scharm}

There is a second choice for which nonzero $\lambda^{''}$ can give rise to the CDF excess. If $\lambda^{''}_{212}$ is turned on, then both $\tilde{s}$ and $\tilde{c}$ can be resonantly produced from $d$-$c$ and $d$-$s$ initial states, respectively. This choice has the advantage that $\lambda^{''}_{212}$ will also induce decays of both the $\tilde{s}$ and the $\tilde{c}$, so no second $\lambda^{''}_{ijk}$ need be turned on. There are several disadvantages however. Since both $\tilde{s}$ and $\tilde{c}$ are now resonantly produced through the same coupling, the UA2 dijet resonance bound on the lighter state (assumed to have a mass of 150 GeV) will directly apply to the production cross section of the heavier state, and therefore on the cross section for the $Wjj$ decay mode. Furthermore, the branching fraction for the W-mode depends on the left-right squark mixing and may be much smaller for the second generation squarks compared to $\tilde{b}$ and $\tilde{t}$. Finally, the production cross section for a $d$-$s$ initial state is lower than that of the $u$-$s$ initial state by a factor of order one. Since this affects the bounds from dijet resonances in the same way as the production cross section however, one can simply compensate by using a larger value of $\lambda^{''}$ as long as one is not already at the upper limit.

\section{Resonant Slepton Production And Decay}
\label{sec:sleptons}

\begin{figure}[t]
\begin{center}
\includegraphics[scale=0.5]{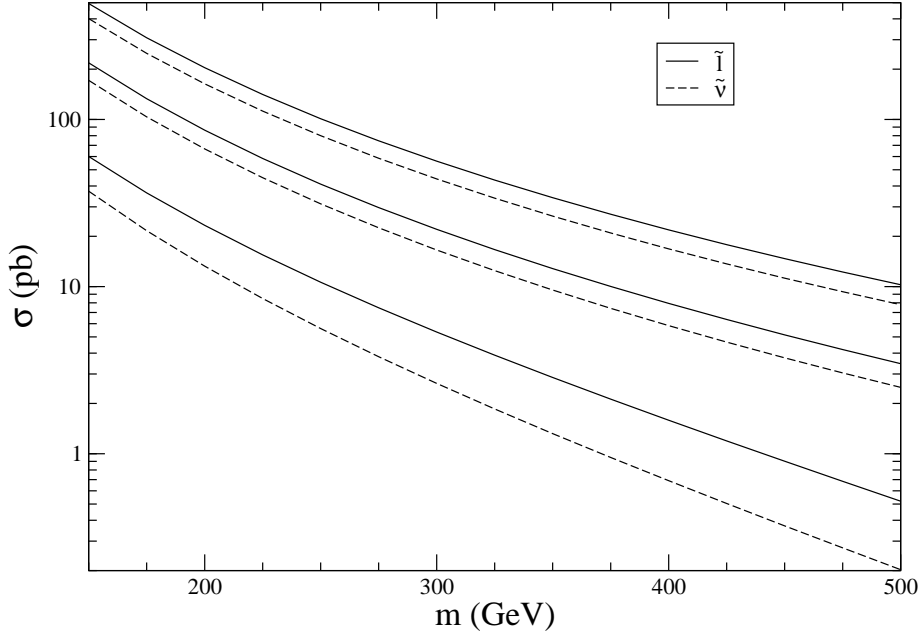}
\end{center}
\caption{Resonant production cross section for a slepton and a sneutrino for $\lambda^{'}_{311}=0.1$. Zero mass mixing is assumed. From bottom to top, the three pairs of curves are for the Tevatron, 7 TeV LHC and 14 TeV LHC respectively.}
\label{fig:sleptonxsec}
\end{figure}

\begin{figure}[t]
\begin{center}
\includegraphics[scale=0.5]{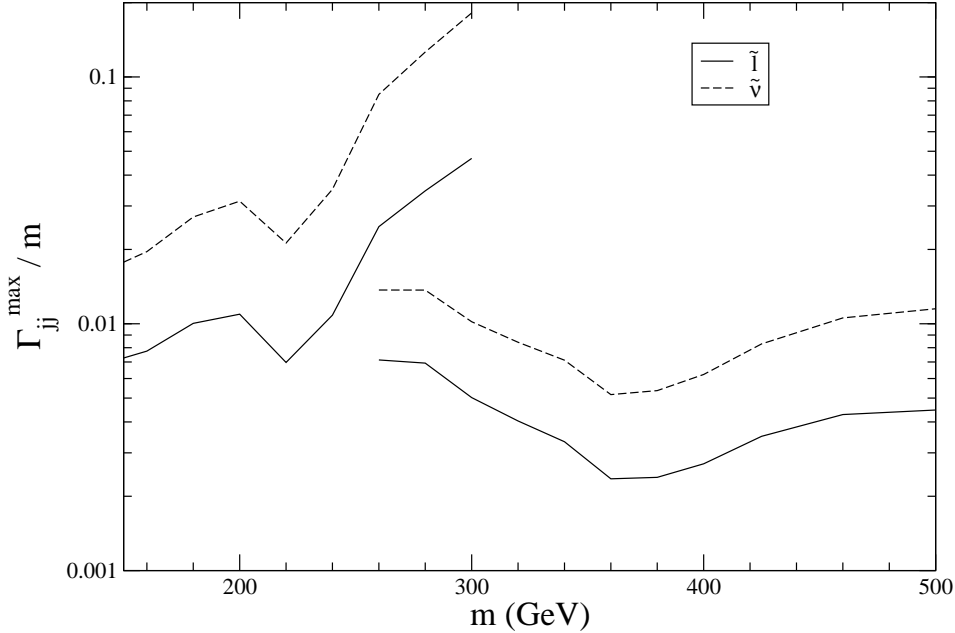}
\end{center}
\caption{The bound on $\Gamma_{jj}$ for resonant slepton and sneutrino production from UA2 and CDF with only $\lambda^{'}_{311}$ turned on. The numbers plotted here are conservative in the sense that we assume perfect acceptance and we neglect left-right mixing. The bound on $\lambda^{'}_{311}$ never goes below 0.3 for the sneutrino and below 0.2 for the slepton.}
\label{fig:sleptonbound}
\end{figure}

$R$-parity violation through $UDD$-type terms is not the only possibility to resonantly produce superpartners at a hadron collider. One can also consider the resonant production of left-handed sleptons through a $LQD$ type coupling. The $\lambda^{'}_{ijk}$ couplings necessary in this scenario are generically more constrained that the $UDD$ couplings \cite{RPVreview}.
While we will not attempt to build a detailed model as in the squark case, we remark that a similar final state can be obtained. Resonant production of a left-handed slepton, with charged-current decay through an off-shell W$^*$ to the sneutrino, which subsequently decays back through the $R$-parity violating coupling, gives an $\ell \bar{\nu}$jj signature. The main difference with a left-handed $SU(2)_L$ doublet only, is that the mass splitting between the slepton and sneutrino arises only from electroweak symmetry breaking and is parametrically small. The leptons in this scenario are then significantly softer than those arising from an on-shell W topology.

The production cross section and the bound from dijet resonance searches for this scenario are plotted in figures \ref{fig:sleptonxsec} and \ref{fig:sleptonbound} respectively. Rather than go through all the details of the phenomenology again, we highlight here the differences compared to the squark case. Since many $\lambda^{'}_{ijk}$ are constrained through lepton flavor violating processes, there are few choices in which couplings can be turned on to match the CDF $Wjj$ observation. One possibility is resonant stau production from a nonzero value of $\lambda^{'}_{311}$. Scenarios of $R$-parity violation with $\tilde{\tau}$ and $\tilde{\nu}_{\tau}$ LSP have been considered in mSUGRA scenarios (see for example \cite{Desch:2010gi}, \cite{Dreiner:2008ca} and references therein). As in the squark case, there is a charge asymmetry in the resonant production of the slepton vs. the anti-slepton. Unlike the squark case however, the origin of this asymmetry is the difference between the $u$ and $d$ valence quark PDF's. Therefore the asymmetries of the signal and the W+jets background will be of the same size. Additionally, unlike in the case of squarks, sleptons are only pair produced through electroweak interactions, so pair production cross sections will be much smaller, and only $q$-$\bar{q}$ initial states can contribute, which is a disadvantage for the LHC compared to the squark case where the $g$-$g$ initial state contributes.

If both right- and left-handed sleptons are light enough to be produced, then the same production and decay topology discussed above can arise with resonant production of the heavier states that charged-current decay through an on-shell W-boson to the lighter states which subsequently decay back to two jets. Since in $LQD$-type $R$-parity violation both resonant production as well as the charged-current coupling involve the left-handed component of sleptons, significant left-right mixing is required in order for both the cross section as well as the W branching fraction to be sizeable in this case. If kinematically open, decay modes with on-shell Higgs and $Z$-bosons also arise in this scenario.

\section{Conclusions}
\label{sec:conclusions}

We have explored a production and decay topology for hadron colliders in which a primary resonance undergoes a charged-current transition to a secondary resonance that subsequently decays to dijets. Supersymmetric scenarios with $R$-parity violation where a light sbottom is resonantly produced and decays to a W-boson and stop, which decays to a pair of jets through the $R$-parity violating coupling, give a concrete realization of this topology. Other possibilities for underlying theoretical frameworks that could give topologies of this type include technicolor \cite{techni} (for reviews of phenomenology see \cite{TC-reviews,Lane:2002sm,Brooijmans:2008se}), two-Higgs-doublet models (for a detailed overview see \cite{Gunion:1989we}), and excited quarks \cite{Terazawa:1981eg,Renard:1982ya,DeRujula:1983ak}. We have argued that with such a topology, the new physics may be observed first in the $Wjj$ channel, possibly consistent with the recent CDF observation in this channel. We have also pointed to other final states that would be associated with this scenario. In particular, for strongly interacting resonances, other important topologies are pair production of the secondary state giving rise to a 4 jet signature, as well as pair production of the primary state giving rise to a W$^+$W$^-$ + 4 jets signature, both with two equal mass dijet resonances within the 4 jets. We argued that due to the smallness of the mass of the secondary state, the former would most effectively be searched for at the Tevatron, while the latter has better prospects at the upgraded LHC. We have outlined how a less minimal supersymmetric model can be set up with all three generations of squarks present, without significantly changing the collider phenomenology, but forbidding additional $R$-parity violating couplings to be induced at loop-level from existing ones.

We have also shown that there is an alternative choice of couplings for the squark scenario, where the phenomenology is focused on the second generation squarks rather than the third. This alternative model needs only one $R$-parity violating coupling to be nonzero, however the branching fraction for the W mode may be small unless there is significant left-right mixing in the squark sector. Finally we have outlined an alternative supersymmetric scenario with the same production and decay topology, but where the new particles are sleptons rather than squarks.
Many additional production channels such as pair production through QCD are absent in this case.

To conclude, we believe that even if the CDF $Wjj$ observation proves not to be due to new physics, that collider signatures of $R$-parity violating supersymmetric models should be further explored by both theorists and experimental searches.

{\bf Note Added:} After this paper was submitted, several developments took place. CDF updated their analysis with a $7.3~{\rm fb}^{-1}$ data set in which the significance of the excess was reported as $4.1~\sigma$ after accounting for systematic uncertainties \cite{CDFupdate}. In this analysis, various other consistency checks were performed on the analysis, such as using different Monte Carlo generators, making the jet cuts inclusive (rather than demanding exactly two jets), varying the jet energy scale and some of the analysis cuts. The excess was reported as being robust under these variations. This is true of the signal in our scenario as well, the effect of these variations to the acceptance of our Monte Carlo sample is mild. There is however a result in the updated CDF analysis that has a larger impact on our scenario, namely the heavy flavor content in the jets accompanying the $W$-boson has been reported quantitatively. This result is difficult to reconcile with one of the two jets being a b-jet in each event, and therefore disfavors our sbottom-stop benchmark. Note however that both the second generation squark benchmark as well as the slepton benchmark which we proposed produce light jets in the final state and are therefore still consistent with the CDF update.

Another dramatic development was the release of a D\O\ search for new physics in the same final state \cite{Abazov:2011af}. This analysis reports that D\O\ does not observe an excess consistent with the CDF result, and for a Gaussian centered at the same mass they obtain a best fit for the cross section of $0.82^{+0.83}_{-0.82}~{\rm pb}$. Considering that the D\O\ analysis uses cuts very similar to the CDF analysis, it is not at the moment clear why the two experiments obtain such drastically different results. While this development casts doubt on the existence of new physics in this final state, it is also possible that there is a signal with a cross section smaller than the one reported by CDF but still consistent with the D\O\ result, for example around $1.5~{\rm pb}$. If that were to be the case, the signal in our scenario could easily have a smaller cross section by using a higher mass for the first resonant squark or slepton, or by having smaller $R$-parity violating couplings. As we remarked at the end of our conclusions, as theorists we believe this topology for the resonant production of supersymmetric particles to be an interesting one that should be studied further even if the eventual resolution of the discrepancy between CDF and D\O\ results disfavors a new physics explanation.

We also wish to mention a variety of other new physics explanations that were proposed to explain the CDF excess after the submission of this paper. These include technicolor models \cite{TCmodels}, new gauge bosons \cite{WZprime}, extended Higgs sectors \cite{2HDMmodels} and other possibilities \cite{othermodels}. Since technicolor is arguably the most commonly associated type of new physics associated with the $Wjj$ final state, we want to point to a few phenomenological differences between our scenario and a technicolor model, which could be used in order to distinguish the two signals. For one, since a techni-pion is expected to couple to SM fermions proportionally to their mass, it is expected to decay preferentially to $b\bar{b}$ pairs. In our scenario, the alternative squark and slepton benchmarks would only give rise to light jets in the final state whereas the sbottom-stop benchmark would produce a single $b$-jet. As mentioned earlier in this note, the CDF update appears to disfavor the existence of $b$-tagged jets in the excess. Another possibility that can distinguish between our scenario and a technicolor model is to look for pair production of the primary resonance. In both our sbottom-stop as well as our alternative squark benchmarks, both the heavier as well as the lighter squarks are expected to be pair produced from QCD processes, giving rise to a $4j$ final state for the lighter squark and a $WW+4j$ final state for the heavier squark, as we mentioned at the end of section \ref{sec:squarks}, whereas techni-rhos and techni-pions in minimal technicolor models do not carry color charge and will not be pair produced in the same way. Note however that pair production is not a good discriminant if the resonantly produced superpartners are sleptons which also do not carry color.

\subsection*{Acknowledgments}
The work of C.K. and S.T. is supported by DOE grant DE-FG02-96ER40959.



\begin{thebibliography}{99}


\bibitem{RPVreview}
  R.~Barbier {\it et al.},
  Phys.\ Rept.\  {\bf 420}, 1 (2005)
  [arXiv:hep-ph/0406039].

\bibitem{SUSYcolliderpheno}
  S.~Dawson, E.~Eichten, C.~Quigg,
  Phys.\ Rev.\  {\bf D31}, 1581 (1985);
  H.~Baer, X.~Tata,
  Cambridge, UK: Univ. Pr. (2006) 537 p;
  K.~Nakamura {\it et al.} [ Particle Data Group Collaboration ],
  J.\ Phys.\ G {\bf G37}, 075021 (2010).

\bibitem{Dimopoulos:1988fr}
  S.~Dimopoulos, R.~Esmailzadeh, L.~J.~Hall and G.~D.~Starkman,
  Phys.\ Rev.\  D {\bf 41}, 2099 (1990).

\bibitem{Dreiner:1991pe}
  H.~K.~Dreiner and G.~G.~Ross,
  Nucl.\ Phys.\  B {\bf 365}, 597 (1991).

\bibitem{Sarid:1999zx}
  U.~Sarid and S.~D.~Thomas,
  Phys.\ Rev.\ Lett.\  {\bf 85}, 1178 (2000)
  [arXiv:hep-ph/9909349].

\bibitem{Sakai:1981pk}
  N.~Sakai and T.~Yanagida,
  Nucl.\ Phys.\  B {\bf 197}, 533 (1982).

\bibitem{Hall:1983id}
  L.~J.~Hall and M.~Suzuki,
  Nucl.\ Phys.\  B {\bf 231}, 419 (1984).

\bibitem{Brahm:1989iy}
  D.~E.~Brahm and L.~J.~Hall,
  Phys.\ Rev.\  D {\bf 40}, 2449 (1989).

\bibitem{Hempfling:1995wj}
  R.~Hempfling,
  Nucl.\ Phys.\  B {\bf 478}, 3 (1996)
  [arXiv:hep-ph/9511288].

\bibitem{Smirnov:1995ey}
  A.~Y.~Smirnov and F.~Vissani,
  Nucl.\ Phys.\  B {\bf 460}, 37 (1996)
  [arXiv:hep-ph/9506416].

\bibitem{Tamvakis:1996xf}
  K.~Tamvakis,
  Phys.\ Lett.\  B {\bf 382}, 251 (1996)
  [arXiv:hep-ph/9604343].

\bibitem{Tamvakis:1996dk}
  K.~Tamvakis,
  Phys.\ Lett.\  B {\bf 383}, 307 (1996)
  [arXiv:hep-ph/9602389].

\bibitem{Barbieri:1997zn}
  R.~Barbieri, A.~Strumia and Z.~Berezhiani,
  Phys.\ Lett.\  B {\bf 407}, 250 (1997)
  [arXiv:hep-ph/9704275].

\bibitem{Giudice:1997wb}
  G.~F.~Giudice and R.~Rattazzi,
  Phys.\ Lett.\  B {\bf 406}, 321 (1997)
  [arXiv:hep-ph/9704339].

\bibitem{Dreiner:1997uz}
  H.~K.~Dreiner,
  arXiv:hep-ph/9707435.

\bibitem{Diaz:1997xc}
  M.~A.~Diaz, J.~C.~Romao and J.~W.~F.~Valle,
  Nucl.\ Phys.\  B {\bf 524}, 23 (1998)
  [arXiv:hep-ph/9706315].

\bibitem{Allanach:2003eb}
  B.~C.~Allanach, A.~Dedes and H.~K.~Dreiner,
  Phys.\ Rev.\  D {\bf 69}, 115002 (2004)
  [Erratum-ibid.\  D {\bf 72}, 079902 (2005)]
  [arXiv:hep-ph/0309196].

\bibitem{Sundrum:2009gv}
  R.~Sundrum,
  JHEP {\bf 1101}, 062 (2011)
  [arXiv:0909.5430 [hep-th]].

\bibitem{Sundrum_private}
  R.~Sundrum, private communication.

\bibitem{CDFanomaly}
  T.~Aaltonen {\it et al.}  [CDF Collaboration],
  arXiv:1104.0699 [hep-ex].

\bibitem{UA2bound}
 UA2 Collaboration,
  Nucl.\ Phys.\ B\ {\bf 400}, 3 (1993)
  ISSN 0550-3213, DOI: 10.1016/0550-3213(93)90395-6.

\bibitem{CDF_dijet}
  T.~Aaltonen {\it et al.}  [CDF Collaboration],
  Phys.\ Rev.\  D {\bf 79}, 112002 (2009)
  [arXiv:0812.4036 [hep-ex]].

\bibitem{Aad:2011aj}
  G.~Aad {\it et al.}  [ATLAS Collaboration],
  arXiv:1103.3864 [hep-ex].

\bibitem{Khachatryan:2010jd}
  V.~Khachatryan {\it et al.}  [CMS Collaboration],
  Phys.\ Rev.\ Lett.\  {\bf 105}, 211801 (2010)
  [arXiv:1010.0203 [hep-ex]].

\bibitem{CTEQ6}
J.~Pumplin, D.~R.~Stump, J.~Huston, H.~L.~Lai, P.~M.~Nadolsky and W.~K.~Tung,
JHEP {\bf 0207} (2002) 012
[arXiv:hep-ph/0201195].

\bibitem{Boos:2001cv}
  E.~Boos {\it et al.},
  arXiv:hep-ph/0109068.

\bibitem{PYTHIA}
  T.~Sjostrand, S.~Mrenna and P.~Z.~Skands,
  ``PYTHIA 6.4 Physics and Manual,''
  JHEP {\bf 0605}, 026 (2006)
  [arXiv:hep-ph/0603175].

\bibitem{PGS}
  J.~Conway {\em et~al.},
  ``{PGS 4: Pretty Good Simulation of high energy collisions},'' 2006,
  {\tt www.physics.ucdavis.edu/$\sim$conway/research/software/pgs/pgs4-general.htm}

\bibitem{calchep}
A.~Pukhov {\it et al.},
arXiv:hep-ph/9908288;
%
A.~Pukhov,
arXiv:hep-ph/0412191.

\bibitem{Choudhury:2005dg}
  D.~Choudhury, M.~Datta and M.~Maity,
  Phys.\ Rev.\  D {\bf 73}, 055013 (2006)
  [arXiv:hep-ph/0508009].

\bibitem{Chivukula:1991zk}
  R.~S.~Chivukula, M.~Golden and E.~H.~Simmons,
  Nucl.\ Phys.\  B {\bf 363}, 83 (1991).

\bibitem{Dobrescu:2007yp}
  B.~A.~Dobrescu, K.~Kong and R.~Mahbubani,
  Phys.\ Lett.\  B {\bf 670}, 119 (2008)
  [arXiv:0709.2378 [hep-ph]].

\bibitem{Kilic:2008pm}
  C.~Kilic, T.~Okui and R.~Sundrum,
  JHEP {\bf 0807}, 038 (2008)
  [arXiv:0802.2568 [hep-ph]].

\bibitem{Kilic:2008ub}
  C.~Kilic, S.~Schumann and M.~Son,
  JHEP {\bf 0904}, 128 (2009)
  [arXiv:0810.5542 [hep-ph]].

\bibitem{Bai:2010dj}
  Y.~Bai and B.~A.~Dobrescu,
  arXiv:1012.5814 [hep-ph].

\bibitem{Desch:2010gi}
  K.~Desch, S.~Fleischmann, P.~Wienemann, H.~K.~Dreiner and S.~Grab,
  Phys.\ Rev.\  D {\bf 83}, 015013 (2011)
  [arXiv:1008.1580 [hep-ph]].

\bibitem{Dreiner:2008ca}
  H.~K.~Dreiner and S.~Grab,
  Phys.\ Lett.\  B {\bf 679}, 45 (2009)
  [arXiv:0811.0200 [hep-ph]].

\bibitem{techni}
  L.~Susskind,
  ``Dynamics Of Spontaneous Symmetry Breaking In The Weinberg-Salam Theory,''
  Phys.\ Rev.\ D {\bf 20}, 2619 (1979);
  S.~Weinberg,
  ``Implications Of Dynamical Symmetry Breaking,''
  Phys.\ Rev.\ D {\bf 13}, 974 (1976);
  S.~Weinberg,
  ``Implications Of Dynamical Symmetry Breaking: An Addendum,''
  Phys.\ Rev.\ D {\bf 19}, 1277 (1979).

\bibitem{TC-reviews}
  E.~Farhi and L.~Susskind,
  ``Technicolor,''
  Phys.\ Rept.\  {\bf 74}, 277 (1981);
  C.~T.~Hill and E.~H.~Simmons,
  ``Strong dynamics and electroweak symmetry breaking,''
  Phys.\ Rept.\  {\bf 381}, 235 (2003)
  [Erratum-ibid.\  {\bf 390}, 553 (2004)]
  [arXiv:hep-ph/0203079].

\bibitem{Lane:2002sm}
  K.~Lane and S.~Mrenna,
  ``The Collider phenomenology of technihadrons in the technicolor straw man model,''
  Phys.\ Rev.\  D {\bf 67}, 115011 (2003)
  [arXiv:hep-ph/0210299].

\bibitem{Brooijmans:2008se}
  G.~H.~Brooijmans {\it et al.},
  ``New Physics at the LHC: A Les Houches Report. Physics at Tev Colliders 2007
  -- New Physics Working Group,''
  arXiv:0802.3715 [hep-ph].

\bibitem{Gunion:1989we}
  J.~F.~Gunion, H.~E.~Haber, G.~L.~Kane and S.~Dawson,
  Front.\ Phys.\  {\bf 80}, 1 (2000).

\bibitem{Terazawa:1981eg}
  H.~Terazawa, M.~Yasue, K.~Akama and M.~Hayashi,
  Phys.\ Lett.\  B {\bf 112}, 387 (1982).

\bibitem{Renard:1982ya}
  F.~M.~Renard,
  Nuovo Cim.\  A {\bf 77}, 1 (1983).

\bibitem{DeRujula:1983ak}
  A.~De Rujula, L.~Maiani and R.~Petronzio,
  Phys.\ Lett.\  B {\bf 140}, 253 (1984).

\bibitem{CDFupdate}
  {\tt http://www-cdf.fnal.gov/physics/ewk/2011/wjj/7\_3.html}

\bibitem{Abazov:2011af}
  V.~M.~Abazov [ D0 Collaboration ],
  Phys.\ Rev.\ Lett.\  {\bf 107}, 011804 (2011).
  [arXiv:1106.1921 [hep-ex]].
  
\bibitem{TCmodels}
  E.~J.~Eichten, K.~Lane, A.~Martin,
  Phys.\ Rev.\ Lett.\  {\bf 106}, 251803 (2011).
  [arXiv:1104.0976 [hep-ph]].

\bibitem{WZprime}
  M.~R.~Buckley, D.~Hooper, J.~Kopp, E.~Neil,
  Phys.\ Rev.\  {\bf D83}, 115013 (2011).
  [arXiv:1103.6035 [hep-ph]];
  F.~Yu,
  Phys.\ Rev.\  {\bf D83}, 094028 (2011).
  [arXiv:1104.0243 [hep-ph]];
  X.~-P.~Wang, Y.~-K.~Wang, B.~Xiao, J.~Xu, S.~-h.~Zhu,
  Phys.\ Rev.\  {\bf D83}, 117701 (2011).
  [arXiv:1104.1161 [hep-ph]];
  K.~Cheung, J.~Song,
  Phys.\ Rev.\ Lett.\  {\bf 106}, 211803 (2011).
  [arXiv:1104.1375 [hep-ph]];
  A.~E.~Nelson, T.~Okui, T.~S.~Roy,
  [arXiv:1104.2030 [hep-ph]];
  S.~Jung, A.~Pierce, J.~D.~Wells,
  [arXiv:1104.3139 [hep-ph]];
  P.~Ko, Y.~Omura, C.~Yu,
  [arXiv:1104.4066 [hep-ph]];
  P.~J.~Fox, J.~Liu, D.~Tucker-Smith, N.~Weiner,
  [arXiv:1104.4127 [hep-ph]];
  D.~-W.~Jung, P.~Ko, J.~S.~Lee,
  [arXiv:1104.4443 [hep-ph]];
  S.~Chang, K.~Y.~Lee, J.~Song,
  [arXiv:1104.4560 [hep-ph]];
  K.~S.~Babu, M.~Frank, S.~K.~Rai,
  [arXiv:1104.4782 [hep-ph]];
  Z.~Liu, P.~Nath, G.~Peim,
  Phys.\ Lett.\  {\bf B701}, 601-604 (2011).
  [arXiv:1105.4371 [hep-ph]];
  J.~L.~Hewett, T.~G.~Rizzo,
  [arXiv:1106.0294 [hep-ph]];
  A.~E.~Faraggi, V.~M.~Mehta,
  [arXiv:1106.5422 [hep-ph]].
  
\bibitem{2HDMmodels}
  Q.~-H.~Cao, M.~Carena, S.~Gori, A.~Menon, P.~Schwaller, C.~E.~M.~Wagner, L.~-T.~Wang,
  [arXiv:1104.4776 [hep-ph]];
  B.~Dutta, S.~Khalil, Y.~Mimura, Q.~Shafi,
  [arXiv:1104.5209 [hep-ph]];
  C.~-H.~Chen, C.~-W.~Chiang, T.~Nomura, Y.~Fusheng,
  [arXiv:1105.2870 [hep-ph]].

\bibitem{othermodels}  
  X.~-P.~Wang, Y.~-K.~Wang, B.~Xiao, J.~Xu, S.~-h.~Zhu,
  Phys.\ Rev.\  {\bf D83}, 115010 (2011).
  [arXiv:1104.1917 [hep-ph]];
  R.~Sato, S.~Shirai, K.~Yonekura,
  Phys.\ Lett.\  {\bf B700}, 122-125 (2011).
  [arXiv:1104.2014 [hep-ph]];
  L.~A.~Anchordoqui, H.~Goldberg, X.~Huang, D.~Lust, T.~R.~Taylor,
  Phys.\ Lett.\  {\bf B701}, 224-228 (2011).
  [arXiv:1104.2302 [hep-ph]];
  B.~A.~Dobrescu, G.~Z.~Krnjaic,
  [arXiv:1104.2893 [hep-ph]];
  H.~B.~Nielsen,
  [arXiv:1104.4642 [hep-ph]];
  B.~Bhattacherjee, S.~Raychaudhuri,
  [arXiv:1104.4749 [hep-ph]];
  G.~Segre, B.~Kayser,
  [arXiv:1105.1808 [hep-ph]];
  T.~Enkhbat, X.~-G.~He, Y.~Mimura, H.~Yokoya,
  [arXiv:1105.2699 [hep-ph]].
  
\end{thebibliography}
\end{document}